\documentclass[twocolumn,aps,pra,showpacs,showkeys,preprintnumbers,amsmath,amssymb]{revtex4}
\usepackage{graphicx}
\usepackage{dcolumn}
\usepackage{bm}

\def\vep{{\varepsilon}}

\def\al{\alpha}
\def\be{\beta}
\def\gam{\gamma}
\def\lam{\lambda}

\def\beq{\begin{equation}}
\def\eeq{\end{equation}}
\def\bea{\begin{eqnarray}}
\def\eea{\end{eqnarray}}
\begin{document}
\title{Transition temperature of the homogeneous and
 dilute Bose gas in D-dimensions}
\author{ Sang-Hoon \surname{Kim}
\footnote{shkim@mmu.ac.kr}}
\affiliation{Division of Liberal Arts and Sciences,
Mokpo National Maritime University, Mokpo 530-729, R.O. Korea}
\begin{abstract}
The phase transition temperature of the homogeneous and dilute Bose
gas in D-dimensions ($2 \le D \le 3$) is calculated by a mean
field-based statistical method. The shift of the phase transition
temperature
 is written up to the leading order as
$\Delta T_c/T_c^0  = c \gam^{\al}$,
where $\gam=n^{1/3}a$.
 We derived Huang's result of the phase
transition temperature  in the generalized dimensions.
We show that $
c(D)$ is positive and $\al(D)=2(D/2-1)^2$ in the short-wavelength
range. The origin of the difference between $\al=1/2$ and  $\al=1$
at D=3 is discussed. The $T_c$ at $D=2$  is calculated in the same
scheme. The result is compared with Fisher and Hohenberg's KT
temperature.
\end{abstract}
\pacs{03.75.Hh, 05.30.Jp, 05.70.Fh, 21.60.Fw} \keywords{homogeneous
and dilute Bose system, phase transition temperature, Bose-Einstein
condensation}
\maketitle

\section{Introduction}

The homogeneous and dilute(HD) Bose gas is a fundamental topic of
physics to which a lot of theoretical efforts have been devoted in
recent years. The shift of the transition temperature($T_c$)
remains an interesting problem to this
day. At the same time, the study of $T_c$ in two-dimensions(2D) has
been a crucial topic because it is directly connected with the
Kosterlitz-Thouless(KT) transition.

 In a homogeneous system, the density in D-dimensions is expressed as
$n(D)=n(1)^D$. The diluteness is expressed by a dimensionless gas
parameter $\gam = n(1)a = n(3)^{1/3}  a $, where $n(1)$ is the
one-dimensional number density and $a$ is the $s$-wave scattering
length. In a repulsive and dilute gas, $\gam$ is positive and much
less than 1. The {\it order of diluteness} can be defined as $|\ln
\gam|$, and  much larger than 1.
The
argument in the paper is valid only for an extremely weakly
interacting Bose system,
 and therefore cannot be applied to a strongly interacting system
such as real liquid $^4$He.

The transition temperature of  the non-interacting homogeneous Bose
gas in D-dimensions is \cite{kim1}
\beq T_c^0(D) = \frac{2 \pi
\hbar^2}{m k_B} \frac{n(2)} {\zeta(\frac{D}{2})^\frac{2}{D}},
\label{1}
\eeq
where  $2 \le D \le 3$  and $\zeta$ is the
Riemann-zeta function. The transition temperature of the HD Bose
system is written as a series form of $\gam$. The leading orders in
3D and 2D are \bea T_c(3) &=& T_c^0(3)(1+ c \gam^\al + \cdots),
\label{5} \\
T_c(2) &=& T_c^0(3)(0+ \frac{c'}{|\ln \gam|^{\al'}} + \cdots).
\label{6}
 \eea
The difficulty in finding the $c$ comes from a
breakdown of perturbation theory at the second order transition
temperature.
 Comparing with Eq. (\ref{5}), the interacting system is
equivalent to using the effective mass
$m^\ast = m/(1+c\gam^\al)$
in the ideal gas expression for $T_c^0(3)$. The
positive $c$ means that the effective mass of the quasiparticle is
reduced and  vice versa for the negative $c$.

Two choices of $\al$ for the leading order have been determined with a
repulsive hard-sphere interaction
$\al=1/2$, suggested by Toyota and Huang  from  mean field-based microscopic
theories \cite{toyota,huang,huang3}. On the other hand, $\al=1$ was
suggested
 by others \cite{stoof,cruz,arnold,kashunikov,baym1}.
 Toyota and Huang reported the coefficient $c$ as
\beq
|c|=\frac{8\sqrt{2\pi}}{3\zeta
\left(\frac{3}{2}\right)^{\frac{2}{3}}}=3.524,
\label{7}
\eeq
but the others claimed that $c$ is on the order of 1.
Nowadays, $\al=1$ and
positive $c$ has been widely accepted since
 numerical simulations \cite{arnold,kashunikov}.

 Fisher and Hohenberg(FH)'s
paper has suggested a KT transition temperature in 2D for
the HD Bose gas from the Bogoliubov theory under the condition that
 $\ln \ln \gam^{-2} \gg 1$.
  It has a double logarithmic
form as \cite{fh}
\beq T_c^{FH}(2) = \frac{2 \pi \hbar^2}{m k_B} \frac{n(2)} {
\ln \ln \gam^{-2} }. \label{10} \eeq At a glance, FH's expression
looks different from the Eq. (\ref{6}).

In this paper, we will calculate the  $\al(D)$ and $c(D)$ of the HD
Bose gas between 2 and 3 dimensions with a semi-classical method. We
will explain why Toyota and Huang's leading order $\al=1/2$ is
different from others  where $\al=1$. The KT transition temperature in 2D
will be obtained under the same scheme and compared with Hohenberg
and Fisher's, too.

\section{Equation of states in $D$-dimensions}

For a momentum-independent interaction, the mean field
contribution to the self-energy $\Sigma(D)$ is $ n(D) U(D) $\cite{zagrebnov},
 where $U(D)$ is the D-dimensional coupling constant.
 Then, the dispersion
relation of the weakly interacting Bose gas in D-dimensions from the
mean field theory is \beq \vep =  \vep_0 + \Sigma(D), \label{20}
\eeq where $\vep_0=p^2/2m$. It corresponds to a short-wavelength
range in the Bogoliubov energy spectrum \cite{zagrebnov}, and it
help us to calculate $T_c$ without any breakdown problem.
In
principle, this scheme draws from Hartree-Fock theory, which assumes that
 the atoms behave as non-interacting bosons in a
self-consistent mean field \cite{giorgini1,dalf}.
We may choose the self-energy as
the exact form $2nU$ instead of $nU$ because the two
mean field contributions to the self-energy from Hartree and Fock
are equal.
However, for convenience we will use Eq. (\ref{20})
because the factor of 2 is
out of focus here.

 The  $U(D)$ is
obtained from a D-dimensional scattering theory. Under the
hard-sphere interaction of the diameter $a$, it is well-known in 3D as
\beq
U(3) = \frac{4 \pi \hbar^2 a}{m}.
\label{50}
\eeq
In 2D, on the other hand,
it has a logarithmic form \cite{kim2,lieb}
\beq U(2) =
\frac{2 \pi \hbar^2}{m}\frac{1} {\ln (1/k a)}
= \frac{4 \pi
\hbar^2}{m}\frac{1} {|\ln \gam^2|}.
 \label{55}
\eeq
However, we had better use the D-dimensional interpolated form of
$U(D)$ between two and three dimensions for general purposes \cite{kleinert}.
\beq
U(D)= \frac{4 \pi^{\frac{D}{2}} \hbar^2 a^{D-2}}
{\Gamma\left(\frac{D}{2}-1\right) m}.
\label{52}
\eeq

 The average occupation
number of the momentum-independent potential
 is written from Eq.
(\ref{20}) as
\bea
n_p &=& \frac{1}{z^{-1}e^{\be (\vep_0(p)+\Sigma)}-1}
\nonumber \\
&=& \frac{1}{z_e^{-1}e^{\be \vep_0(p)}-1}, \label{70} \eea where
$\be=1/k_B T$. The $z$ is the  fugacity given by $z=e^{\be \mu}$ and
1 below $T_c$. The $\mu$ is the chemical potential, which is $0$
below the $T_c$ and negative above the $T_c$. The $z_e$ is the
effective fugacity given by \beq z_e = e^{-\be(|\mu|+\Sigma)}\equiv
e^{\beta \mu_e}. \label{80} \eeq Note that  $0 < z_e < 1$.  The
effective chemical potential $\mu_e$ is written in the same way : $\mu_e =
\mu - \Sigma$. Therefore, the interacting system has the same
formula as the ideal system except for the effective chemical
potential and the effective fugacity.

Let us define the ratio between the effective
fugacity and the fugacity  as
 \beq
\eta = \frac{z_e}{z} = e^{- \be \Sigma }.
\\ \label{85}
\eeq Note that $0< \eta_c<1$ and 1 for the ideal system. As $T
\rightarrow T_c$, then $z_e \rightarrow e^{ -\be_c \Sigma} \equiv
\eta_c$.
 When the
interaction is very weak, the phase transition is described by  the
variable :
 $\eta_c \simeq 1 - \be_c \Sigma$.
Therefore, it is clear that even for a
momentum-independent potential,  it shifts the $T_c$.

From Eq. (\ref{70}) we have the D-dimensional equation of states
with $z_e$ as \cite{huang2}
\beq n(D)
\lam^D = g_{\frac{D}{2}}(z_e).
\label{90}
\eeq
$\lam=\sqrt{2\pi\hbar^2/m k_B T}$ is the thermal
wavelength,
and $g_s(z)=\sum_{l=1}^{\infty} z^l /l^s$ is the Bose
function.
 In this way we can write the transition temperature in
D-dimensions, $2 \le D \le 3$,  of the interacting system as \beq
T_c(D) = \frac{2 \pi \hbar^2}{m k_B} \frac{n(2)}
{g_{\frac{D}{2}}(\eta_c)^\frac{2}{D}}. \label{100} \eeq
From Eq.
(\ref{90}) the relation of the transition temperature between ideal
and interacting systems is written as
\beq
\left( T_c^0
\right)^\frac{D}{2} g_{\frac{D}{2}}(1) = T_c^{\frac{D}{2}}
g_{\frac{D}{2}}(\eta_c).
\label{95}
\eeq
As $D$ approaches 2, the above quantity
approaches  a finite value, $2 \pi \hbar^2 n(2)/m k_B$,
from Eq. (\ref{1}). It is about 5.92K for liquid $^4$He.

\section{$T_c$  between 2 and 3 dimensions}

From Eq. (\ref{95}),  it is necessary to expand the $g_{\frac{D}{2}}(\eta_c) $
in an analytic form  to find the transition temperature.
Since $\be_c \Sigma $,
the
interaction term at $T_c$, is much less than 1,
we use the expansion of the
Bose functions for $D>2$ \cite{robinson}. \beq g_{\frac{D}{2}} (e^{
-\be_c \Sigma }) \simeq \zeta\left(\frac{D}{2} \right) +
\Gamma\left(1-\frac{D}{2}\right) (\be_c \Sigma)^{\frac{D}{2} -1}
+{\cal O}(\be_c \Sigma). \label{110} \eeq
Substituting Eq. (\ref{110}) into Eq. (\ref{95}), \bea T_c^{
\frac{D}{2} } &=& \left( T_c^0 \right)^\frac{D}{2} \frac{ \zeta(
\frac{D}{2} ) } {g_{\frac{D}{2}}(\eta_c)}
\nonumber \\
& = & \left( T_c^0 \right)^\frac{D}{2} \left[ 1+ \frac{
\Gamma(1-\frac{D}{2} ) }{ \zeta(\frac{D}{2} ) }(\be_c
\Sigma)^{\frac{D}{2} -1} \right]^{-1}. \label{120} \eea Up to the
linear term, we have \beq T_c   \simeq T^0_c \left[ 1
-\frac{2}{D}\frac{ \Gamma(1-\frac{D}{2} ) } {\zeta(\frac{D}{2} ) }
(\be_c^0 \Sigma)^{\frac{D}{2}-1}  \right]. \label{130} \eeq
From Eqs. (\ref{1}) and (\ref{20})
\bea
\be_c^0(D) \Sigma(D)
 &=& \frac{m \zeta\left(\frac{D}{2} \right)^{\frac{2}{D}}  n(1)^{D-2} U(D)}
{2\pi\hbar^2}
\nonumber \\
& = & \frac{2 \pi^{\frac{D}{2}-1} \zeta\left(\frac{D}{2} \right)^{\frac{2}{D}} }
{\Gamma\left( \frac{D}{2}-1\right) }
\gam^{D-2}.
\label{140}
\eea
Note that
$\be_c^0(3) \Sigma(3)= 2 \zeta\left(\frac{3}{2}\right)^{\frac{2}{3}}\gam$.
 The leading order $\alpha$ of the shift of the chemical potential $\Sigma(3)/k_B$ is 1,
but it
definitely cannot be the shift of  $T_c(3)$.

\begin{figure}
\includegraphics{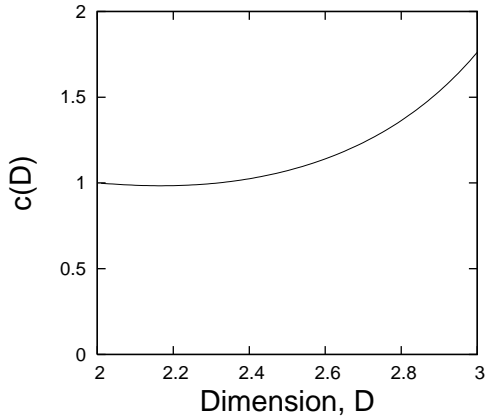}\\
\caption{ $c(D)$ as a function of dimensions.}
\end{figure}

Substituting Eq. (\ref{140}) into Eq. (\ref{130}), we obtain the
D-dimensional $T_c$ as
\beq
T_c(D) = T^0_c(3) \left[ 1-\frac{2^{\frac{D}{2}}
\pi^{\left(\frac{D}{2}-1\right)^2} \Gamma\left(1- \frac{D}{2}\right)
} { D \zeta\left(\frac{D}{2} \right)^{\frac{2}{D}}
\Gamma\left(\frac{D}{2}-1 \right)^{\frac{D}{2}-1}
} \gam^{ 2\left(
\frac{D}{2}-1\right)^2 } \right].
\label{150}
\eeq
Therefore, we
have the two coefficients of the $T_c$ shift in D-dimensions as
\beq
c(D) = -\frac{2^{\frac{D}{2}} \pi^{\left(\frac{D}{2}-1\right)^2}
\Gamma\left(1- \frac{D}{2}\right) } { D\zeta\left(\frac{D}{2}
\right)^{\frac{2}{D}} \Gamma\left(\frac{D}{2}-1
\right)^{\frac{D}{2}-1} } \label{152} \eeq and \beq \al(D)=2\left(
\frac{D}{2}-1\right)^2 , \label{153}
 \eeq
where $2 <D \le 3$. Note that $\al=0$ in 2D as the Eq. (\ref{6}).

There has been a long dispute about the sign of the $c$. From Eq.
(\ref{152}) we see that it is decided by -$\Gamma(1-\frac{D}{2})$.
Therefore, the $c$ is positive at $D > 2$.
It is plotted in FIG. 1.
However, note that the minimum of $c(D)$ is not at $D=2$ but at $D=2.17$.
 If $D=3$, we obtain
$c(3)=4\sqrt{2\pi}/3\zeta(3/2)^{2/3}=1.762$
and
$\al=1/2$.
Note that
$\Gamma(-\frac{1}{2})=-2\sqrt{\pi}$.
 The $c(3)$ is exactly  half of Toyota
and Huang's microscopic results in Eq. (\ref{7}) in that they  chose
$2nU$ instead of $nU$ as the self-energy.
The positive $c$ means that
the BEC temperature, $T_c$, is different from the lambda transition
temperature, $T_\lambda$, and there possibly exists a pseudogap-like region
between $T_c$ and $T_\lambda$ in superfluidity.

\section{$T_c$ in Two-dimensions}

\begin{figure}
\includegraphics{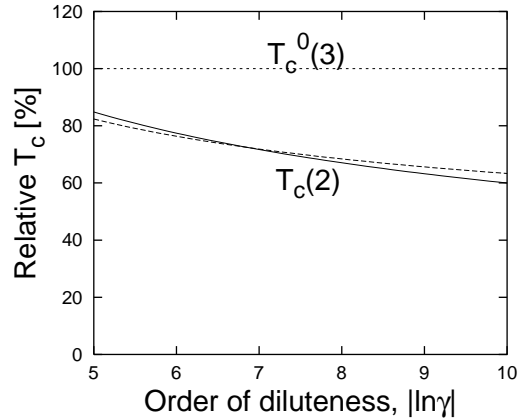}\\
\caption{ The relative transition temperature in 2D compared with
that of the 3D ideal gas. The solid line is our $T_c(2)$ and the
dashed line is $T_c^{FH}(2)$ by Fisher and Hohenberg.}
\end{figure}

For an ideal Bose system, there is no condensation because a
long-wavelength phase fluctuation destroys the long-range order. On
the other hand, in an interacting system, the interparticle
interactions drive a phase transition to a condensate state and
then, there exists a nonzero condensate density above a specific
temperature $T_c^{KT}(2)$.

We can also obtain the transition temperature in 2D from Eq.
(\ref{100}) as
 \beq T_c(2) = \frac{2\pi \hbar^2}{m k_B}
\frac{n(2)}{g_{1}(\eta_c)}. \label{160} \eeq The convergence of
$g_1(\eta_c)$ is extremely slow. Therefore, we need to use another
method for the limiting value. The Euler's constant  $C=0.5772...$
is defined as \beq \lim_{s \rightarrow 1} \left[ \zeta(s) -
\frac{1}{s-1} \right] =C. \label{170} \eeq It is effective up to the
first order even when $\lim_{x\rightarrow 1}g_1(x)$. Then, the
limiting behavior of the $g_1(\eta_c)$ when $\eta_c $ is close to 1
is \beq
 g_1(\eta_c) \simeq
\frac{1}{1-\eta_c} = \frac{1}{\be_c(2) \Sigma(2)}.
\label{180}
\eeq
Substituting Eq. (\ref{180}) into Eq. (\ref{160}),
\beq
T_c(2)=\frac{2 \pi \hbar^2 n(2)^2 U(2) }{m k_B^2 T_c(2) }.
\label{190}
\eeq

Substituting Eq. (\ref{55}) into Eq. (\ref{190}),
 we obtain the 2D transition temperature
as
\beq
T_c(2)=\frac{2 \pi \hbar^2 }{m k_B}
\frac{n(2)}{\sqrt{|\ln
\gam|}}.
\label{200}
 \eeq
The leading orders in Eq. (\ref{6}) are
obtained as $ \al' = \frac{1}{2}, $ and
$ c' = \zeta\left(\frac{3}{2}\right)^{\frac{2}{3}}=1.897. $
 $T_c(2)$ is
shown in FIG. 2, and compared with $T_c^{FH}(3)$ and $T_c^0(3)$ in
the range of interest.

FH follows the approximation for $T_c^{FH}(2)$ under the {\it
ansatz} : $\mu \sim n/|\ln(a^2 \mu)|$ \cite{fh}. Then, the
transition temperature in 2D is proportional to the following :
 \beq
T_c^{FH}(2) \sim \frac{n(2)}{|\ln ka|} \sim \frac{n(2)}{\ln (\mu
a^2)} \sim \frac{n(2)}{\ln \ln \gam^{-2}}. \label{220} \eeq We think
the {\it ansatz} is not necessary, because FH's transition
temperature is nothing but $T_c^{FH}(2) = \Sigma(2)/2 k_B$ or just
the chemical potential in 2D. Note that the factor of half comes
from choosing the self-energy $nU$  instead of $2nU$. This happens
because the clean form of $U(2)$ in Eq. (\ref{55}) was not known at
that time.

\section{Summary}

We derived the transition temperature of the HD Bose gas between 2
and 3 dimensions in the  short-wavelength range using a
semi-classical method. The two coefficients of the leading order of
$T_c(D)$ ($\al$ and  $c$), where $2 < D \le 3$, are obtained as a
function of the dimensions.
 Toyota and Huang's results  are just the
special case of our calculation at $D=3$.
It was
obtained not because the mean field theory does not work, but because
the long-wavelength limit is not included. Therefore, for an
extremely low density and at the short-wavelength range,
we may accept Toyota and Huang's
results of $\al=1/2$ \cite{huang,huang3}.

The KT transition temperature in 2D in the same condition  is
obtained. The two coefficients of the leading order of $T_c(2)$
($\al'$ and  $c'$) are obtained, too. We show that  Fisher and
Hohenberg's KT transition temperature is just the chemical potential
in 2D. The leading orders of the $T_c$ in 3D and 2D are obtained as
$\al=\al'=1/2$ in the short-wavelength range, but the reliable
results of $\al=\al'=1$ are expected if we include the
long-wavelength limit.

\begin{acknowledgments}
The author acknowledges useful discussions with K. Huang.
\end{acknowledgments}

{}
\end{document}